\documentstyle[sprocl]{article}

\def\ga{\gamma}
\def\de{\delta}
\def\ep{\epsilon}

\def\ka{\kappa}
\def\la{\lambda}

\def\si{\sigma}

\def\ps{\psi}
\def\om{\omega}

\def\cl{{\mathcal L}}

\def\mn{{\mu\nu}}

\def\half{{\textstyle{1\over 2}}}

\def\frac#1#2{{\textstyle{{#1}\over {#2}}}}

\def\lsim{\mathrel{\rlap{\lower4pt\hbox{\hskip1pt$\sim$}}
    \raise1pt\hbox{$<$}}}
\def\gsim{\mathrel{\rlap{\lower4pt\hbox{\hskip1pt$\sim$}}
    \raise1pt\hbox{$>$}}}
\def\sqr#1#2{{\vcenter{\vbox{\hrule height.#2pt
         \hbox{\vrule width.#2pt height#1pt \kern#1pt
         \vrule width.#2pt}
         \hrule height.#2pt}}}}

\def\prt{\partial}

\def\etal{{\it et al.}}

\newcommand{\beq}{\begin{equation}}
\newcommand{\eeq}{\end{equation}}
\newcommand{\bea}{\begin{eqnarray}}
\newcommand{\eea}{\end{eqnarray}}
\newcommand{\bit}{\begin{itemize}}
\newcommand{\eit}{\end{itemize}}

\begin{document}

\title{QED Tests of Lorentz Symmetry}

\author{R.\ Bluhm}

\address{Physics Department, \\
Colby College, \\ 
Waterville, ME 04901, USA\\ 
E-mail: rtbluhm@colby.edu}

\maketitle

\abstracts{
A status report is given of some recent theoretical and experimental
investigations looking for signals of Lorentz violation in QED.
Experiments with light, charged particles, and atoms have exceptional 
sensitivity to small shifts in energy caused by Lorentz violation,
including effects that could originate from new physics 
at the Planck scale.}

\section{Introduction}

Lorentz symmetry is a fundamental feature of relativity theory.
In special relativity,
it is a global symmetry relating the laws of
physics in different inertial frames under boosts and rotations.
It is also linked by a general theorem to the combined 
discrete symmetry CPT formed from the product of charge conjugation C,
parity P, and time reversal T.
In general relativity, 
Lorentz symmetry becomes a local symmetry that relates 
the physics in different freely falling frames in 
a gravitational field.

The Standard Model (SM) of particle physics does not include
gravity as a fundamental interaction at the quantum level.
It is therefore expected that the SM and gravity 
will merge in the context of a fundamental unified theory.
The relevant energy scale for quantum gravity is the Planck scale 
$M_{\rm Pl} = \sqrt{\hbar c/G} \simeq 10^{19}$ GeV.
Much current work in theoretical high-energy physics is
aimed at finding a unified fundamental theory that describes
physical interactions at the Planck scale.
Promising candidates include string theory,
D-branes, and theories of quantum gravity.
Many of these include effects that violate
assumptions of the SM,
including higher dimensions of spacetime,
unusual geometries, nonpointlike interactions,
and new forms of symmetry breaking.

In particular,
it is possible that small violations of Lorentz
symmetry might occur in theories of quantum gravity.
For example,
it is known that there are mechanisms in string theory that can
lead to spontaneous violations of Lorentz and CPT symmetry.\cite{kskp}
This is due to certain types of interactions in string theory
among Lorentz-tensor fields that can destabilize the vacuum
and generate nonzero vacuum expectation values for Lorentz tensors.
These vacuum expectation values fill the true vacuum and cause 
spontaneous Lorentz violation.
It is also known that geometries with noncommutative
coordinates can arise naturally in string theory
and that Lorentz violation is intrinsic to noncommutative field
theories.\cite{chlkO01}

One method of searching for signals of Planck-scale physics is
to look for highly suppressed effects involving inverse
powers of the Planck scale.
In this approach, Lorentz violation becomes an ideal
signal since all of the interactions in the SM
preserve Lorentz symmetry and therefore no conventional
signal could mimic the effects of Lorentz violation.
To observe a signal of Lorentz violation experimentally,
one needs to perform experiments with exceptional sensitivity.
Experiments in QED systems provide many of the best
oppportunities for testing Lorentz symmetry.

One example is provided by measurements of photons that have
traveled over cosmological distances.
Any small phase effect would be amplified during
the long transit time.
Other examples with photons include high-precision 
laboratory-based experiments with resonance cavities.
Experiments in atomic physics can also be performed at 
low energy with extremely high precision. 
For example, some atomic experiments are routinely sensitive
to small frequency shifts at the level of 1 mHz or less.
Interpreting this as being due to an energy shift expressed in GeV,
it would correspond to a sensitivity of approximately
$4 \times 10^{-27}$ GeV.
Such a value is well within the range of energy one
might associate with suppression factors originating
from the Planck scale.

The main focus of this work is to investigate tests of Lorentz and 
CPT symmetry performed in QED systems.
The general goals are to analyze the sensitivity
of QED systems to possible Lorentz and CPT violation,
to uncover possible new signals that can be tested in experiments,
and to express experimental sensitivities in the context of a common 
framework that permits comparisons across different experiments.
To this end,
we use the Standard-Model Extension (SME) 
as our theoretical framework.\cite{sme}
The SME permits detailed investigations of Lorentz and CPT tests
in all particle sectors of the SM.
Our analysis here focuses on the QED sector of the SME. 
This is presented in the following section and is then used
to examine a number of experiments involving photons,
trapped particles, atomic clocks, muons, and a spin-polarized pendulum.
Additional details about many of these experiments can be found as 
well in several of the other articles in this volume.

\section{QED Sector of the SME}

The subset of the SME lagrangian relevant to experiments in QED systems
can be written as
\beq
{\mathcal L}_{\rm QED} = {\mathcal L}_0 + {\mathcal L}_{\rm int} 
\quad .
\label{lag}
\eeq
The lagrangian ${\mathcal L}_0$ contains the usual Lorentz-invariant
terms in QED that describe photons, massive charged fermions, 
and their conventional couplings.
If we restrict our investigation to the remormalizable
and gauge-invariant terms in the SME in flat spacetime,
then the Lorentz-violating part of the lagrangian is given by\cite{note1}
\bea
{\mathcal L}_{\rm int} &=& - a_\mu \bar \ps \ga^\mu \ps
- b_\mu \bar \ps \ga _5 \ga^\mu \ps
+ ic_{\mu \nu} \bar \ps \ga^\mu D^\nu \ps 
\nonumber \\
&& \quad\quad
+ id_{\mu \nu} \bar \ps \ga_5 \ga^\mu D^\nu \ps 
- \half H_{\mu \nu} \bar \ps \si^{\mu \nu} \ps 
\nonumber \\
&&
-\frac14 (k_F)_{\ka\la\mu\nu} F^{\ka\la}F^{\mu\nu} 
+ \frac 12 (k_{AF})^\ka \ep_{\ka\la\mu\nu} A^{\la} F^{\mu\nu} 
\quad .
\label{qedsme}
\eea
Here,
natural units with $\hbar = c = 1$ are used,
and $i D_\mu \equiv i \partial_\mu - q A_\mu$.
The terms with coefficients $a_\mu$, $b_\mu$ and $(k_{AF})_\mu$
violate CPT,
while those with
$H_{\mu \nu}$, $c_{\mu \nu}$, $d_{\mu \nu}$, and $(k_F)_{\ka\la\mu\nu}$
preserve CPT.
All seven terms break Lorentz symmetry.

Sensitivities to Lorentz and CPT violation can be
expressed in terms of the SME coefficients.
This provides a straightforward way of making comparisons across different
types of experiments.
Each different particle sector in the QED extension
has an independent set of Lorentz-violating coefficients.
These are distinguished using superscript labels.
A thorough investigation of Lorentz and CPT violation
requires looking at as many different particle sectors as possible.

\section{Photon Experiments}

The relevant part of the lagrangian for a freely propagating
photon in the presence of Lorentz violation is given by
\beq
\cl = -\frac14 F_{\mu\nu}F^{\mu\nu}
-\frac14 (k_F)_{\ka\la\mu\nu}
     F^{\ka\la}F^{\mu\nu}\ 
+ \frac12 (k_{AF})^\ka \ep_{\ka\la\mu\nu} A^{\la} F^{\mu\nu} ,
\eeq
where $F_\mn$ is the field strength,
$F_\mn \equiv \prt_\mu A_\nu -\prt_\nu A_\mu$.

The CPT-odd term with coefficient $k_{AF}$ has been investigated
extensively both theoretically and experimentally.\cite{cfj,jk}
It is found theoretically that this term leads to negative-energy contributions 
and is a potential source of instability.  
One solution is to set $k_{AF} \approx 0$,
which is consistent with radiative corrections in the SME.
Stringent experimental constraints consistent with $k_{AF} \approx 0$ 
have also been determined by studying the polarization of radiation
from distant radio galaxies.\cite{cfj}
In the following,
we will therefore ignore the effects of $k_{AF}$.

The CPT-even term with coefficients $k_{F}$ have been
investigated more recently.\cite{km}
It provides positive-energy contributions.
The set of coefficients $k_{F}$ has 19 independent components.
It is useful to make a decomposition of these in terms of a new set:
$\tilde\ka_{e+}$, $\tilde\ka_{e-}$,
$\tilde\ka_{o+}$, $\tilde\ka_{o-}$, and $\tilde\ka_{\rm tr}$.  
Here, $\tilde\ka_{e+}$, $\tilde\ka_{e-}$, and $\tilde\ka_{o-}$
are $3\times3$ traceless symmetric matrices 
(with 5 independent components each),
while $\tilde\ka_{o+}$ is a
$3\times3$ antisymmetric matrix 
(with 3 independent components),
and the remaining coefficient $\tilde\ka_{\rm tr}$
is the only rotationally invariant component.

The lagrangian in terms of this
decomposition becomes 
\bea
\cl &=& \half [(1+\tilde\ka_{\rm tr})\vec E^2
-(1-\tilde\ka_{\rm tr})\vec B^2]
+\half \vec E\cdot(\tilde\ka_{e+}
+\tilde\ka_{e-})\cdot\vec E
\nonumber \\
&&
-\half\vec B\cdot(\tilde\ka_{e+}
-\tilde\ka_{e-})\cdot\vec B 
+\vec E\cdot(\tilde\ka_{o+}
+\tilde\ka_{o-})\cdot\vec B\ 
\quad .
\label{modL}
\eea
Here, $\vec E$ and $\vec B$ are
the usual electric and magnetic fields.

The equations of motion following from this lagrangian 
give rise to modifications of Maxwell's equations,
which have been explored in several recent astrophysical 
and laboratory experiments.
The ten coefficients $\tilde\ka_{e+}$ and $\tilde\ka_{o-}$ lead to
birefrigence of light.
Spectropolarimetry of light from distant galaxies leads to
bounds on these parameters of order $2 \times 10^{-32}$.\cite{km}
Seven of the eight coefficients $\tilde\ka_{e-}$ and $\tilde\ka_{o+}$ are bounded
in experiments using optical and microwave cavities.\cite{photonsexpts}
Sensitivities on the order of $\tilde\ka_{e-} \lsim 10^{-15}$
and $\tilde\ka_{o+} \lsim 10^{-11}$ have been attained,
and it is expected that future experiments in boosted frames will be sensitive
to the remaining two parameters as well.

\section{Atomic Experiments}

In recent years,
a number of atomic experiments have been performed which
have very sharp sensitivity to Lorentz and CPT violation.
Bounds from these experiments can be expressed in terms of the coefficients
$a_\mu$, $b_\mu$, $c_{\mu \nu}$, $d_{\mu \nu}$, and $H_{\mu \nu}$
in the QED sector of the SME.
Comparisons across different types of experiments can then be made which avoid 
the problems that can arise when different physical quantities
($g$ factors, charge-to-mass ratios, masses, frequencies, etc.)
are used in different experiments.
In the following,
a number of atomic experiments involving the
proton, neutron, electron, and muon are examined.

\subsection{Penning-Trap Experiments}

Two recent sets of experiments with electrons and positrons in Penning traps 
provide sharp tests of Lorentz and CPT symmetry.\cite{bkr9798}
Both involve measurements of the
anomaly frequency $\om_a$ and the cyclotron frequency $\om_c$.

The first consists of a reanalysis by Dehmelt's group of existing data 
for electrons and positrons in a Penning trap.\cite{dehmelt99} 
The signal involves looking for an instantaneous difference in the
anomaly frequencies of electrons and positrons,
which can be nonzero when Lorentz and CPT symmetry are broken.
In contrast the instantaneous cyclotron frequencies remain approximately equal
at leading order in the Lorentz-violation corrections.
Dehmelt's original measurements of $g-2$
did not involve looking for possible instantaneous variations in $\om_a$.
Instead,
the ratio $\om_a/\om_c$ was computed using averaged values.
It is important to realize that the Lorentz-violating corrections to the 
anomaly frequency $\om_a$ can occur even though the $g$ factor remains unchanged.
The new analysis looks for an
instantaneous difference in the electron and positron anomaly frequencies.
A bound on this difference can be expressed in terms of the parameter $b^e_3$,
which is the component of $b^e_\mu$ along the quantization
axis in the laboratory frame.
It is given as $|b^e_3| \lsim 3 \times 10^{-25}$ GeV.

A second signal for Lorentz and CPT violation in the electron
sector has been obtained using data for the electron alone.\cite{mittleman99}
In this case,
the idea is that the Lorentz-violating interactions depend on
the orientation of the quantization axis in the laboratory frame,
which changes as the Earth turns on its axis.
As a result,
both the cyclotron and anomaly frequencies have small corrections which
cause them to exhibit sidereal time variations.
These variations can be measured using electrons alone,
eliminating the need for comparison with positrons.
The bounds in this case are given with respect to a
nonrotating coordinate frame such as celestial equatorial coordinates.
The interactions involve a combination of laboratory-frame components
that couple to the spin of the electron.
This combination is denoted using tildes as
$\tilde b_3^e  \equiv b_3^e - m d_{30}^e - H_{12}^e$.
When expressed in terms of components $X$, $Y$, $Z$ in the nonrotating
frame,
the obtained bound is
$|\tilde b_J^e| \lsim 5 \times 10^{-25} {\rm GeV}$ for $J=X,Y$.

\subsection{Clock-Comparison Experiments}

The Hughes-Drever experiments are 
classic tests of Lorentz invariance.\cite{kl99,cctests}
There have been a number of different types of these
experiments performed over the years,
with steady improvements in their sensitivity.
They all involve making high-precision comparisons of 
atomic clock signals as the Earth rotates.
The clock frequencies are typically hyperfine or Zeeman transitions.
Many of the sharpest Lorentz bounds for the proton, neutron, and electron
stem from atomic clock-comparison experiments.
For example,
Bear {\it et al.} use a two-species noble-gas maser to
test for Lorentz and CPT violation in the neutron sector.\cite{dualmaser}
They obtain a bound
$|\tilde b_J^n| \lsim 10^{-31} {\rm GeV}$ for $J=X,Y$,
which is currently the best bound for the neutron sector.

Note that these Earth-based laboratory experiments are not 
sensitive to Lorentz-violation coefficients along the $J=Z$
direction parallel to Earth's rotation axis.  
They also neglect the velocity effects due to Earth's 
motion around the sun,
which would lead to bounds on the timelike components along $J=T$.
These limitations can be overcome by performing experiments in space\cite{space}
or by using a rotation platform.
The earth's motion can also be taken into account.
A recent boosted-frame analysis of the dual noble-gas maser
experiment yields bounds on the order of $10^{-27}$ GeV
on many boost-dependent SME coefficients for the neutron
that were previously unbounded.\cite{cane}
 
It should also be pointed out that certain assumptions about the nuclear
configurations must be made to obtain bounds in
clock-comparison experiments.
For this reason,
these bounds should be viewed as good to within about
an order of magnitude.
To obtain cleaner bounds it is necessary to consider
simpler atoms or to perform more sophisticated nuclear modeling.

\subsection{Hydrogen-Antihydrogen Experiments}

The simplest atom in the periodic table is hydrogen,
and the simplest antiatom is antihydrogen.
There are three experiments underway at CERN that 
can perform high-precision Lorentz and CPT tests in antihydrogen.
Two of the experiments (ATRAP and ATHENA) intend to
make high-precision spectroscopic measurements of the 1S-2S
transitions in hydrogen and antihydrogen.
These are forbidden (two-photon) transitions that have a relative linewidth
of approximately $10^{-15}$.
The ultimate goal is to measure the line center of this
transition to a part in $10^3$ yielding a frequency comparison
between hydrogen and antihydrogen at a level of $10^{-18}$.
An analysis of the 1S-2S transition in the context of the
SME shows that the magnetic field plays an important role
in the attainable sensitivity to Lorentz and CPT violation.\cite{bkr99}
For instance,
in free hydrogen in the absence of a magnetic field,
the 1S and 2S levels are shifted by equal amounts at leading order.
As a result,
in free H or $\bar {\rm H}$ there are no leading-order corrections 
to the 1S-2S transition frequency.
In a magnetic trap,
however,
there are fields that can mix the spin states in the
four different hyperfine levels.
Since the Lorentz-violating interactions depend on the spin orientation,
there will be leading-order sensitivity
to Lorentz and CPT violation in comparisons of 1S-2S transitions in
trapped hydrogen and antihydrogen.
At the same time,
however,
these transitions are field-dependent,
which creates additional experimental challenges 
that would need to be overcome.

An alternative to 1S-2S transitions is to consider the sensitivity
to Lorentz violation in ground-state Zeeman hyperfine transitions.
It is found that there are leading-order corrections in these levels
in both hydrogen and antihydrogen.\cite{bkr99}
The ASACUSA group at CERN is planning to measure the Zeeman hyperfine 
transitions in antihydrogen.
Such measurements will provide a direct CPT test.
Experiments with hydrogen alone have been performed using a maser.\cite{Hmaser}
They attain exceptionally sharp sensitivity to Lorentz and CPT 
violation in the electron and proton sectors of the SME.
These experiments use a double-resonance technique that does
not depend on there being a field-independent point for the transistion.
The sensitivity for the proton attained in these experiments 
is $|\tilde b_J^p| \lsim 10^{-27}$ GeV.
Due to the simplicity of hydrogen,
this is an extremely clean bound and is currently the most stringent test
of Lorentz and CPT violation for the proton.

\subsection{Spin-Polarized Matter}

Experiments at the University of Washington using a spin-polarized
torsion pendulum\cite{eotwash}
are able to achieve very high sensitivity to
Lorentz violation in the electron sector.\cite{bk00}
The sensitivity arises because the pendulum has a huge
number of aligned electron spins but a negligible magnetic field. 
The pendulum is built out of a stack of toroidal magnets,
which in one version of the experiment achieved
a net electron spin $S \simeq 8 \times 10^{22}$.
The apparatus is suspended on a rotating turntable and  
the time variations of the twisting pendulum are measured.
An analysis of this system shows that in addition to a signal having the
period of the rotating turntable,
the effects due to Lorentz and CPT violation also cause additional
time variations with a sidereal period caused by the rotation
of the Earth.
The group at the University of Washington has analyzed their data
and find that thay have sensitivity to the electron coefficients
at the levels of $|\tilde b_J^e| \lsim 10^{-29}$ GeV for $J=X,Y$ and
$|\tilde b_Z^e| \lsim 10^{-28}$ GeV.\cite{eotwash}
These are currently the best Lorentz and CPT bounds for the electron.
More recently,
a new pendulum has been built,
and it is expected that improved sensitivities will
be attained.

\subsection{Muon Experiments}

Muons are second-generation leptons.
Lorentz tests performed with muons are therefore independent
of the tests involving electrons.
There are two main classes of experiments with muons
that have been conducted recently.
These are experiments with muonium\cite{muonium99}
and $g-2$ experiments with muons at Brookhaven.
In muonium,
measurements of the frequencies
of ground-state Zeeman hyperfine transitions
in a strong magnetic field have the greatest sensitivity
to Lorentz and CPT violation.
An analysis searching for sidereal time variations
in these transitions was able to attain
sensitivity to Lorentz violation at the level of 
$| \tilde b^\mu_J| \le 2 \times 10^{-23}$ GeV.
At Brookhaven,
relativistic $g-2$ experiments with positive muons
have been conducted using muons with boost parameter $\de = 29.3$.
An analysis of the obtained data as a test of Lorentz symmetry 
is still forthcoming.
We estimate that a sensitivity to Lorentz violation is
possible in these experiments at a level of $10^{-25}$ GeV.\cite{bkl00}

\section{Conclusions}

Experiments in QED systems continue to provide many of the
sharpest tests of Lorentz and CPT symmetry.
In recent years,
a number of new astrophysical and laboratory 
tests have been performed that have lead
to substantially improved sensitivities for the photon.
Similarly,
atomic experimentalists continue to find ways of improving 
the sensitivity to Lorentz violation in many of the matter sectors of the SME.
In particular,
experiments in boosted frames are providing sensitivity
to many of the previously unprobed SME coefficients.
All of the bounds obtained are within the range of sensitivity associated
with suppression factors arising from the Planck scale.
The coming years are likely to remain productive.
QED experiments will continue to provide
increasingly sharp new tests of Lorentz and CPT symmetry.


\begin{thebibliography}{99}
\bibitem{note1}
For simplicity,
the additional terms involving parameters $e_\mu$, $f_\mu$,
and $g_{\la\mu\nu}$ are ignored.
Gauge invariance excludes these terms in the SME;
however,
they can arise effectively due to strong binding in a nucleus.
A more general discussion includes an investigation 
of these terms.\cite{kl99}
\end{thebibliography}
\end{document}